\documentclass{aa}
\usepackage{amsmath}
\usepackage{graphicx}
\usepackage{txfonts}
\usepackage{longtable}
\begin{document}
   \title{Mechansim of very high energy radiation in BL Lacertae object 3C 66A}

   \author{Jianping Yang
          \inst{1,2}
          \and
          Jiancheng Wang\inst{1}
          }

   \institute{National Astronomical Observatories, Yunnan Observatory, Chinese Academy
of Sciences,  Kunming 650011, China\\
              \email{yangjp@mail.ynao.ac.cn}
         \and
             Yunnan Agricultural University,
Kunming 650201, China
             }

   \date{Received July 31, 2009; accepted  November 16, 2009}


  \abstract
   {}
   {Our goal is to understand the nature of blazars and the mechanisms for the generation of high-energy $\gamma$-rays, through the
investigation of the blazar 3C 66A.}
   {We model the high energy spectrum of 3C 66A, which has been observed recently
with the Fermi-LAT and VERITAS telescope. The spectrum has a hard
change from the energy range of 0.2-100 GeV to 200-500 GeV in recent
almost contemporaneous observations of two telescopes. }
   {
The de-absorbed VERITAS spectrum greatly depends on the redshift,
which is highly uncertain. If z=0.444 is adopted, we are able to use
the SSC model to produce the Fermi-LAT component and the EC model to
the VERITAS component. However, if z=0.1, the intrinsic VERITAS
spectrum will be softer, there will be a smooth link between the
Fermi-LAT and VERITAS spectra which can be explained using a SSC
model. }
 {}
\authorrunning {J. Yang \& J. Wang}
\titlerunning {high-energy emissions of the 3C 66A}

   \maketitle
%

\section{INTRODUCTION}
Blazars are a peculiar class of active galactic nuclei (AGN) and
their jets point at small angles with respect to our line of sight.
Many of them have been observed at all wavelengths, from radio to
very high energy (VHE) $\gamma$-rays. Their spectral energy
distribution (SED) consists of two bumps which are attributed to the
synchrotron and the inverse Compton (IC) emission of
ultrarelativistic particles. The different soft photon sources
deduce synchrotron self-Compton (SSC) and external Compton (EC)
models to produce high energy emission. In the SSC model, the soft
photons are provided by the synchrotron emission of the same
electrons (\cite{Marscher80}; \cite{Ghisellini89};
\cite{Marscher96}). However in the EC model, the soft photons mostly
come from the outside of the jet, such as outer disk, broad-line
region (BLR) clouds (etc. \cite{Dermer93}; \cite{Sikora94}).

In 3C 66A, \cite{Miller78} have given the redshift z=0.444 by a weak
Mg II emission line detection, but it is very uncertain
(\cite{Bramel05}). When 3C 66A locates at z=0.444, its TeV photons
will suffer the strong pair production absorption of the
Extragalactic Background light (EBL). After corrected by the EBL
absorption, TeV emission presents an inverted intrinsic spectrum
(see the Acciari et al. (2009) Fig. 2, the de-absorbed photon
spectral index is calculated to be $1.1 \pm 0.4$). In this paper, we
take z = 0.444 to analyze TeV emissive mechanism, and will discuss
the behavior of the de-absorbed VERITAS spectrum in different
redshifts. Generally, 3C 66A is classified as a low-frequency peaked
BL Lac object (LBL). The peak of the low-frequency component of LBLs
usually lies in the IR or optical regime, whereas the peak of
high-energy component locates at several GeV. The luminosity of
$\gamma$-ray is typically comparable to or slightly higher than the
luminosity of the synchrotron component. Such as, \cite{Joshi07}
have argued that for 3C 66A the peak of low-frequency component
locates at the optical regime, the peak of high-frequency component
reaches multi MeV-GeV range. However, \cite{Perri03} have revealed
that the synchrotron peak locates in between $10^{15}$ and $10^{16}$
Hz, then 3C 66A is classified as an intermediate-frequency peaked BL
Lac (IBL). From the X-ray spectrum with the photon spectral index
$\Gamma \sim 2.5$ (Bottcher et al. 2005; Donato et al. 2005;
Foschini et al. 2006), which might be the tail of the synchrotron
emission, 3C 66A is considered as an IBL in this paper. 3C 66A is
observed in radio, IR, optical, X-rays and $\gamma$-rays, and shows
large luminosity variations. As described in \cite{Bottcher05}, the
object exhibits several outbursts in the optical band and the
variations of $\Delta$m $\sim$ 0.3-0.5 over a timescale of several
days. Until now, the majority of BL Lacs detected at VHE (very high
energy: E $>$ 100 GeV) are HBLs (high-frequency peaked BL Lacs).
Only IBL W Comae (\cite{Acciari08}), LBL BL Lacertae
(\cite{Albert07}) and 3C 279 (\cite{Albert08}) display the potential
to enlarge the extragalactic TeV source. For 3C 66A, the Crimean
Astrophysical Observatory have reported a 5.1$\sigma$ detection
above 900 GeV(Stepanyan et al. 2002).

Recently, VERITAS have carried out 14 hours' observations for 3C 66A
from September 2007 through January 2008 (hereafter, the 2007-2008
season), and from September through November 2008 (hereafter, the
2008-2009 season) a further 46 hours' data have been taken
(\cite{Acciari09}). Due to the limited spatial resolution of
Cherenkov telescopes it is difficult to accurately identify the
emission region. The radio galaxy 3C 66B lies in the same view field
of 3C 66A at a separation of 0.12$^{\circ}$ and is also a plausible
source of VHE radiation (\cite{Tavecchio08}). The recent detection
by MAGIC favored 3C 66B as VHE source and excluded 3C 66A at an 85\%
confidence level (\cite{Aliu09}). However , VERITAS have found that
3C 66A lies 0.01$^{\circ}$ from the fit position while 3C 66B lies
0.13$^{\circ}$ away, and 3C 66A is VHE source. If 3C 66A has a
redshift of z = 0.444, its de-absorbed spectral index is 1.1$\pm
$0.4 showing very hard intrinsic spectrum (\cite{Acciari09}). In
first three months, the Fermi-LAT Gamma-ray Space Telescope have
observed 3C 66A (\cite{Abdo09}), almost at contemporaneous
observation with VERITAS in the 2008-2009 season. However, very soft
spectrum with the spectral index of 1.97$\pm $0.04 appears in the
Fermi-LAT observing energy range. The $\gamma$-ray spectrum suddenly
hardens from 0.2-100 GeV to 200-500 GeV and challenges the one-zone
homogeneous SSC model.

In Section 2 we present the jet models for application to 3C 66A. We
use the observed data to constrain the model parameters in Section
3. We finish with discussions and conclusions in Section 4.
Throughout this paper, we use a soft cosmology with a deceleration
factor $q_0 = 0.5$ and a Hubble constant $H_0 = 75 km s^{-1}
Mpc^{-1}$.

\section{THE MODELS}

We use a public model of \cite{Georganopoulos07} to describe the
observed spectrum of 3C 66A. The emission region is assumed to be a
sphere (blob) with radius $R$, permeated by a homogeneous magnetic
field $B$. The blob moves with bulk Lorentz factor $\Gamma$ through
an external photon field with a black body spectrum, at an angle
$\theta$ (in this work $\theta \sim 1/\Gamma$ is assumed) with
respect to the line of sight, and has a Doppler factor $\delta \sim
\Gamma$. The relativistic electrons are continuously injected into
the blob at a rate $Q_e(\gamma)= Q_0 \gamma^{-s}$ [$cm^{-3} s^{-1}$]
between $\gamma_{\rm min}$ and $\gamma_{\rm max}$, where
$\gamma_{min}$ is the minimum Lorentz factor of the injected
electrons and should not be confused with the minimum Lorentz factor
of the emitting particles. The injection correspond to a luminosity
$L_{\rm inj}$,
\begin{equation}
{Q_0 =} \frac{L_{\rm inj}(2-s)}{V m_e c^2(\gamma_{max}^{2-s}-\gamma_{min}^{2-s})} (s \neq  2); \\
{Q_0 =} \frac{L_{\rm inj}}{V m_e c^2  ln
\frac{\gamma_{max}}{\gamma_{min}}} (s=2).
\end{equation}
\makeatletter\let\@alph\@@@alph\makeatother Where $V=4\pi R^3/3 $ is
the volume of the blob. Injected electrons might obtain energy from
acceleration processes before into the emission region, such as
diffusive shock acceleration, second order Fermi acceleration and
gradual shear acceleration. Detailed discussions about the
acceleration processes are beyond the scope of this paper. The
electrons suffer synchrotron and inverse Compton losses and
eventually escape from the emission region.

The time-dependent evolution of the electron population $N (\gamma,
t)$ \textbf{[$cm^{-3}$]} inside the emission region is governed by,

\begin{equation}
\label{1} {\partial N (\gamma, t) \over \partial t} = -{\partial
\over \partial \gamma} \left[\left({d\gamma \over dt}\right)_{loss}
N (\gamma, t)\right] + Q_e (\gamma) - \frac{N (\gamma, t)}{t_{esc}}.
\end{equation}
Here, $(d\gamma/dt)_{\rm loss}$ is the radiative energy loss rate,
due to synchrotron, SSC and/or EC emission. $t_{\rm esc}$ is the
electron escape timescale which is several light crossing times, in
this work 5 R/c is adopted.

The energy loss rates of electrons caused by the synchrotron are
given by: $\dot\gamma_{\rm s}=\frac{4\sigma_{\rm T}}{3m_{\rm e} c}
\gamma^2 U_{\rm B}$, where $U_{\rm B} = B^2/(8\pi)$ is the magnetic
energy density. The energy loss rates of inverse Compton emission,
$\dot\gamma_{\rm IC}$, have excellent analytical expressions for the
Thomson regime x $\ll $ 1 and for the deep Klein-Nishina (KN) regime
x $\gg $ 1, but have not ones for the middle regime of $x = \epsilon
\gamma $ ($\epsilon$ is the energy of the incoming photon in units
of the electron rest mass). To overcome this,
\cite{Georganopoulos07} modified an analytical approximation used by
\cite{Moderski05}. $\dot \gamma_{\rm IC}= \frac{4\sigma_{\rm
T}}{3m_{\rm e} c} \gamma^2 U_{\rm r} F_{KN}$, where $U_{\rm r}$ is
the energy density of seed photons including synchrotron photons and
external photons (such as the reprocessed photons by Broad Line
Region (BLR) (\cite{Sikora94})), $F_{KN}$ is given by
\cite{Moderski05}. In this paper, the latter photons are assumed to
be a blackbody radiation with peak frequency $\nu_{\rm ext}$ and
energy density $U_{\rm ext}$ (all seen in the observer frame). For
the beaming of the EC emission, we use the recipe of
\cite{Georganopoulos01}.

\section{\textbf{MODELING} PARAMETERS}

First of all, we use the observed quantities to estimate the
physical parameters in the blob, and then use these values to
reproduce the observed SED. \cite{mastichiadiskirk97} and
\cite{Konopelko03} have estimated the parameter relations of the
inverse Compton scattering in Klein-Nishina regime in a homogeneous
SSC scenario. However, \cite{Paggi09} have found that the simple
relations of parameters in Thomson regime satisfy the observations
of LBL or IBL sources, and that the relations in the extreme KN
limit is not suited to HBLs, indicating that the inverse Compton
scattering for HBLs just borders the KN regime. For 3C 66A, the
Fermi-LAT spectra showing flat and rising shapes and higher energy
observations of VERITAS both indicate that the IC scattering of
Fermi-LAT spectra do unlikely enter into the Klein-Nishina regime.
Therefore, we assume that the Fermi-LAT spectrum just cover the peak
of SSC emission, and use the parameter relations in Thomson regime
to rudely estimate the parameters. In fact, we consider the
Klein-Nishina effect in producing the high energy spectra.

The size of the emitting region, R, can be constrained by
measurements of variability timesscales, i.e. R$\leq \delta c
t_{var}/(1+z)$. The observations of VERITAS have shown the
variability to be the time-scale of days (\cite{Acciari09}). The
multiwave campaign of 3C 66A by the Whole Earth Blazar Telescope
(WEBT) in 2007-2008 have observed several bright flares on time
scales of $\sim$ 10 days (Bottcher et al. 2005). These observations
suggest the size of the emitting region to be about $10^{16}cm$ if
we assume $\delta$ is order of 10. From the peak frequencies of
synchrotron and inverse Compton radiation, we can estimate the
$\gamma_{peak}$ (i.e. the Lorentz factor of the electrons emitting
at the peaks of the synchrotron and SSC components) follow as
(\cite{Tavecchio98}):
\begin{equation}
\nu_s=\frac{4}{3}\nu_L \gamma^2_{peak}\delta /(1+z),
\end{equation}

\begin{equation}
\nu_{SSC} \simeq  \frac{4}{3}\nu_s \gamma^2_{peak},
\end{equation}
and we obtain
\begin{equation}
\gamma_{peak} \simeq  (\frac{3\nu_{SSC}}{4\nu_s })^{1/2},
\end{equation}
and
\begin{equation}
\delta B \simeq (1+z) \frac{\nu_s^2}{\nu_{SSC} \cdot (e / 2\pi m_e
c)}, \label{deta_B}
\end{equation}
where, $\nu_L=e B/ 2\pi m_e c \simeq 2.8 \times 10^6 B$ Hz. From the
equation
\begin{equation} \frac{L_{SSC}}{L_s}= \frac{L_{SSC}^{'}}{L_s^{'}} =\frac{\dot\gamma_{\rm
peak,SSC}}{\dot\gamma_{\rm peak,s}} \simeq \frac{U^{'}_{s}}{U^{'}_B}
\simeq \frac{L_s}{\delta^4 4\pi R^2 c}\frac{8\pi}{B^2} =
\frac{2L_s}{\delta^4 R^2 B^2 c} ,
\end{equation}
we have
\begin{equation}
\delta^4 B^2 \simeq  \frac{2L_s^2}{L_{SSC} R^2 c} \simeq \frac{8 \pi
d_l^2 \cdot (\nu_s F_{\nu,s})^2 }{(\nu_{SSC} F_{\nu,SSC}) \cdot R^2
c} . \label{deta4_B2}
\end{equation}
where $L_{s}$ and $L_{SSC}$ are the observed total luminosity of the
synchrotron peak and SSC peak, $L_{SSC}^{'}$ and $L_{s}^{'}$ are the
luminosity in the comoving frame, and $\dot\gamma_{\rm peak,s}$ and
$\dot\gamma_{\rm peak,SSC}$ denote the synchrotron and SSC cooling
rates of the electrons, $U^{'}_{s}$ and $U^{'}_B$ are the comoving
energy densities of synchrotron photons and magnetic fields, the
$d_l$ is the luminosity distance, $F_{\nu,s}$ and $F_{\nu,SSC}$ are
the energy flux at peaks. Using the equation (\ref{deta_B}) and
(\ref{deta4_B2}) and $ R \approx \delta c t_{var}/(1+z)$ (here,
$t_{var} \approx 2$ days is assumed), we are able to derive the
$\delta$ and B from the $\nu_s$, $\nu_{SSC}$, $\nu_s F_{\nu,s}$, and
$\nu_{SSC} F_{\nu,SSC}$. We take $\nu_s\approx 4\times 10^{15}$ Hz
(Perri et al. 2003), $\nu_s F_{\nu,s}\approx 7\times 10^{-11} erg
\cdot cm^{-2} s^{-1}$ estimated from historical data,
$\nu_{SSC}\approx 1.21\times 10^{25}$ Hz which is the middle band of
Fermi-LAT, $\nu_{SSC}F_{\nu,SSC}\approx 4.9\times 10^{-11} erg\cdot
cm^{-2} s^{-1}$ which is an approximative value observed by the
Fermi-LAT, and z=0.444. We then get $\delta\sim 29$ and B$\sim 0.025
G$. $\delta$ can also be estimated by other methods. The apparent
speed $\beta_{app}$=12.1c has been presented by \cite{Bottcher05}.
Modeling the non-simultaneous SED of 3C 66A, Ghisellini et al.
(1998) suggested $\Gamma \sim $14, which is a typical value of
blazars. \cite{Jorstad01} favored a high superluminal speed up to
$\beta_{app} \approx 27c$ and got $\Gamma \geq 27$ or $\delta \sim
30$. However, it is noted that we use the flux of the Fermi-LAT band
to estimate the parameters, if the modeling spectrum includes the
VERITAS spectrum, the estimated parameters will be modified. In
fact, modeling quasi-simultaneous observations of Fermi-LAT and
VERITAS obtains $\delta=35$ (see the Table 1.).

The steady state distribution of emitting electrons is given by:
\begin{equation}
N(\gamma) = \frac{\int [Q_e(\gamma)- N({\gamma})/t_{esc}]
d\gamma}{\dot\gamma }.
\end{equation}
What is the criterion for steady state? The code of
\cite{Georganopoulos07} firstly takes a time step equal to the
cooling time of the lowest energy electrons under the synchrotron
and external Compton losses, and then calculates $N(\gamma)$ to a
steady state where the injected electrons have fully been cooled,
through an adaptive time step.

We emphasize the selection of the spectral index $s$ of injected
electrons. Based on the X-ray photon spectral index of
\textbf{$\Gamma \approx 2.5$} (Bottcher et al. 2005;
\cite{Donato05}; \cite{Foschini06}), we deduce the spectral index of
emitting electrons to be 4. Assuming the observed X-rays to be from
synchrotron emission of the cooled electrons, the spectral index of
the injected electrons is given by $s=3$. \cite{Celotti08} adopted
$s=3.6$ to model the SED of 3C 66A. Therefore, we use $s \sim 3$ to
model the observed data. The particle injection, radiative cooling,
and escape from the emission region might yields a temporary
quasi-equilibrium state described by a broken power-law. The balance
between escape and radiative cooling will lead to a break in the
equilibrium particle distribution at a break Lorentz factor
$\gamma_b$, where $t_{esc} = t_{cool}(\gamma_b)$. The cooling time
scale is evaluated taking into account synchrotron, SSC and EC
cooling. Depending on whether $\gamma_b$ is greater than or less
than $\gamma_{min}$, the system will be in the slow cooling or fast
cooling regime. In the fast cooling regime ($\gamma_b <
\gamma_{min}$), the equilibrium distribution will be a broken
power-law with $N(\gamma) \propto \gamma ^{-2}$ for $\gamma_b <
\gamma <  \gamma_{min}$ and $N(\gamma) \propto \gamma^{-(s+1)}$ for
$\gamma_{min} < \gamma < \gamma_{max}$. In the slow cooling regime
($\gamma_b > \gamma_{min}$), the equilibrium distribution will be
$N(\gamma) \propto \gamma ^{-2}$ for $ \gamma <  \gamma_{min}$,
$N(\gamma) \propto \gamma^{-s}$ for $\gamma_{min} < \gamma <
\gamma_{b}$ and $N(\gamma) \propto \gamma^{-(s+1)}$ for $\gamma_{b}
< \gamma < \gamma_{max}$. Since for thin synchrotron emission the
energy spectral index is related to that of the emitting electrons
as $\alpha = (q-1)/2$ , where q is the spectral index of the
emitting electrons distribution, and it is immediate to see that the
peak in the $\nu F_{\nu}$ spectrum occurs where q=3. When $s>3$ (in
this work s=3.4 and 3.3 are adopted), $\gamma_{min}$ (the minimum
Lorentz factor of the injected electrons) almost corresponds to
$\gamma_{peak}$
(\cite{Ghisellini98}; \cite{Ghisellini02}). $\gamma_{max}$ presents
the balance between the acceleration and cooling and has small
impact upon the SED, it is usually taken to be $10^5-10^7$ (Inoue \&
Takahara 1996).

Considering the EC emission of electrons, we need to estimate
$U_{ext}$ and $\nu_{ext}$ by mainly considering the soft photons
reprocessed by the BLR. $\nu_{ext}$ is usually considered to be
around optical-UV wavebands, we let $\nu_{ext}=2.5 \times 10^{15}$
Hz. The $L_{ext}$ of FSRQ is easily estimated from emission line
spectra or UV-excesses, while it is difficult to estimate the
$L_{ext}$ of BL Lacs. Assuming the luminosity of accretion disk to
be $L_D=1\times 10^{45}erg \cdot s^{-1}$, which is larger than the
jet luminosity and does not produce a blue bump in the simulated SED
(Joshi \& Bottcher 2007), and taking $R_{ext}\sim 10^{17}$ cm
(\cite{Tavecchio08}), we get the upper-limit $U_{ext,u}=2.65\times
10^{-2}erg \cdot cm^{-3}$ assuming the reprocessing efficiency of
the BLR to be 0.1. In this work, we estimate $U_{ext}$ through
modeling the spectrum under the condition of $U_{ext} \leq
U_{ext,u}$. Taking \textbf{$U_{ext}=3.5\times 10^{-6}erg \cdot
cm^{-3}$}, we can already reproduce the observed data of VERITAS.
Therefore the soft photons reprocessed by the BLR can provide the EC
emission of electrons to produce high energy radiation.

\begin{figure}
   \centering
   \includegraphics[width=10cm]{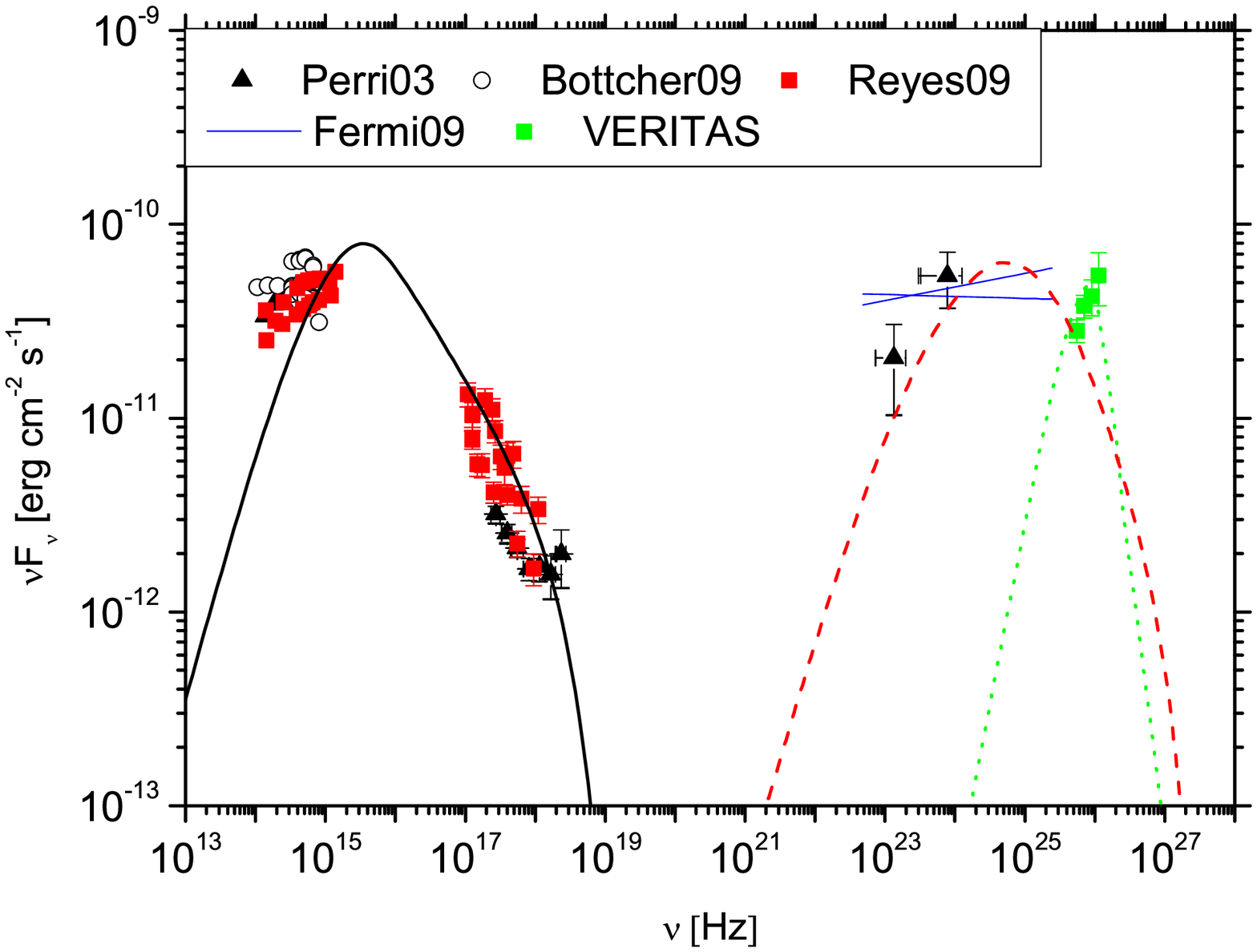}
      \caption{Modeling SED of 3C 66A with the SSC+EC model. The black
triangles are the data of \cite{Perri03}, the open circles come from
\cite{Bottcher09}, the cross blue bow-tie show the Fermi-LAT
spectrum, the red squares denote the data of \cite{Reyes09}, and the
green squares represent the VERITAS spectrum corrected by EBL
absorption according to \cite{Franceschini08}. We use the
synchrotron model to model the lower energy part (black solid line)
, the SSC model to reproduce the Fermi-LAT spectrum (red dash line),
and the EC model to model the VERITAS spectrum (green dot line).
              }
         \label{fig.1}
   \end{figure}

In the Fig.~\ref{fig.1}, we present the modeling results for the
observed data of 3C 66A. The solid triangles are the data of
\cite{Perri03}, the open circles come from \cite{Bottcher09}, the
cross blue bow-tie show the Fermi-LAT data, and the green squares
denote the VERITAS results corrected by EBL absorption according to
\cite{Franceschini08}. In the figure, we include a
quasi-simultaneous data including the near-infrared, optical,
UV-Optical, and X-ray observations (Reyes et al. 2009). We use the
synchrotron emission model to model the lower energy part (black
solid line), and use the SSC model to reproduce the spectrum
observed by the Fermi-LAT (red dash line). Particularly we use the
EC model to model the VERITAS spectrum (green dot line). It is
indicated that the harden spectrum from the Fermi-LAT to VERITAS
energy ranges could exhibit an EC spectrum.

\section{DISCUSSIONS AND CONCLUSIONS}

\begin{figure}
   \centering
   \includegraphics[width=10cm]{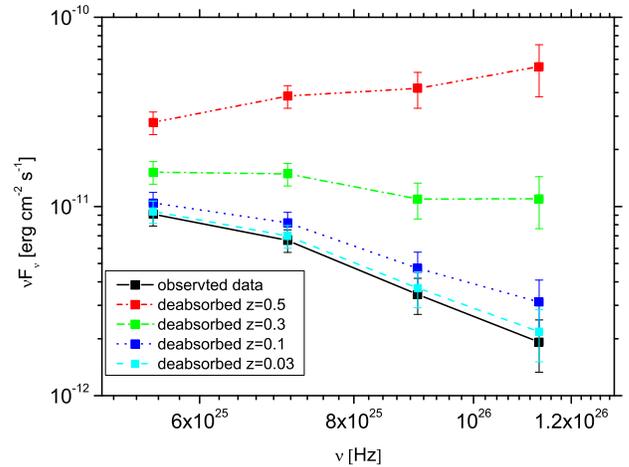}
      \caption{Intrinsic spectra of VERITAS observations corrected by EBL
absorption according to \cite{Franceschini08}
      model in different redshifts, such as z=0.03 (cyan dash line), 0.1 (blue dot line), 0.3 (green dash-dot line), 0.5
(red dash-dot-dot line).
              }
         \label{fig.2}
   \end{figure}

\begin{figure}
   \centering
   \includegraphics[width=10cm]{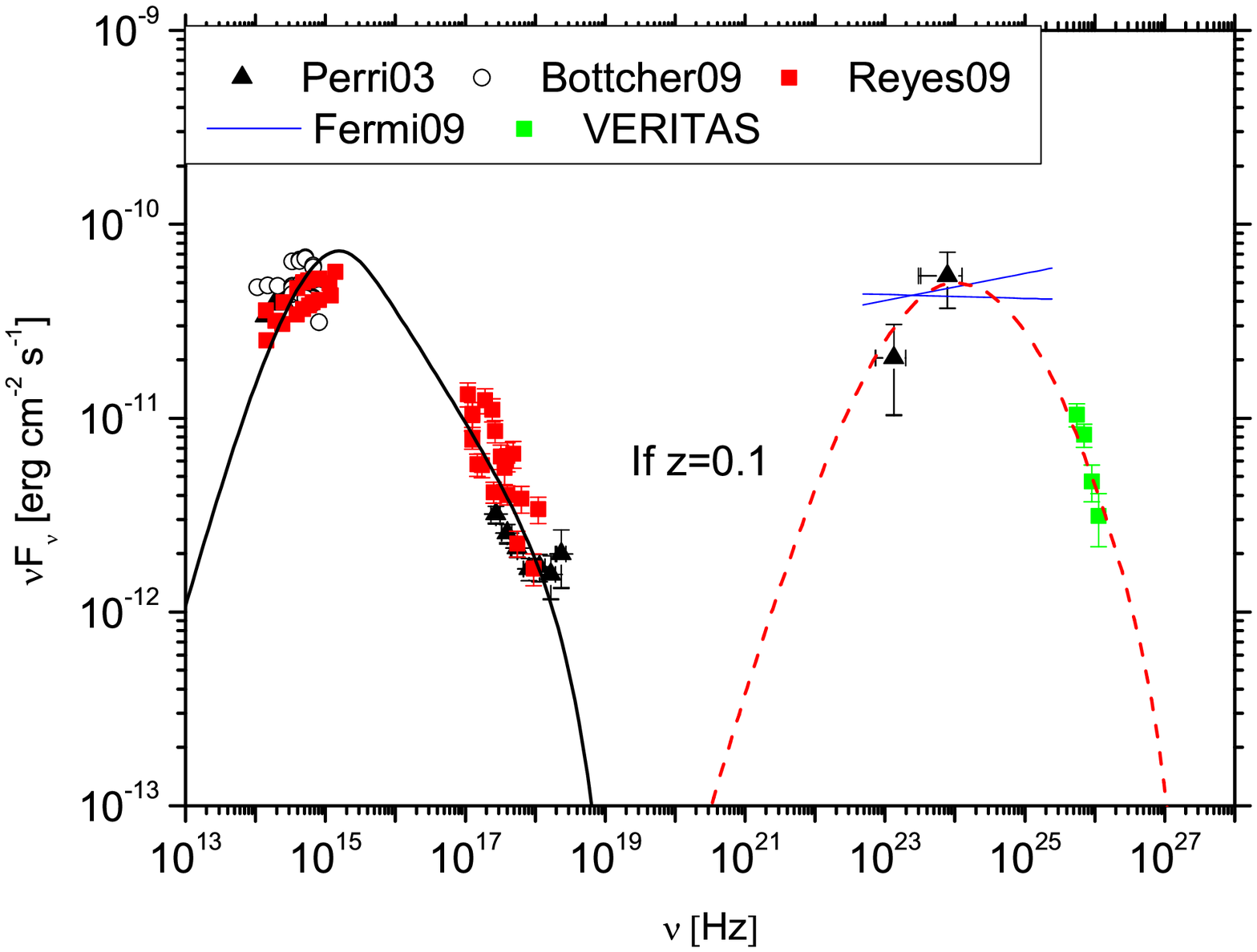}
      \caption{Modeling SED of 3C 66A with the SSC model. The black
triangles are the data of \cite{Perri03}, the open circles come from
\cite{Bottcher09}, the cross blue bow-tie show the Fermi-LAT
spectrum, the red squares present the data of \cite{Reyes09}, and
the green squares denote the VERITAS spectrum corrected by EBL
absorption according to \cite{Franceschini08}. We use the
synchrotron model to model the lower energy part (black solid line)
and the SSC model to reproduce the Fermi-LAT and de-absorbed VERITAS
spectra assuming z=0.1 (red dash line).
              }
         \label{fig.3}
   \end{figure}

The redshift of 3C 66A has an uncertain value (\cite{Bramel05}), and
is usually adopted as z = 0.444. However the redshift is crucial in
constructing intrinsic high energy spectrum because of the EBL
absorption (\cite{Hauser01}). This absorption decreases the observed
flux and softens the observed spectrum. If the redshift is less than
0.444, such as just $\geq$ 0.096 suggested by \cite{Finke08}, the
intrinsic spectrum in the VERITAS energy range will be softer. There
will be a smooth link spectrum between the Fermi-LAT and the VERITAS
energy ranges. In the Fig.~\ref{fig.2}, we generate the intrinsic
spectra of VERITAS observations corrected by EBL absorption
according to \cite{Franceschini08} model, assuming the source to be
at the different redshifts z=0.03, 0.1, 0.3, and 0.5. It is shown
that the de-absorbed spectrum strongly depends on the redshift. When
z=0.3, the de-absorbed spectrum has a little inverted, but it
becomes an inverted spectrum in z=0.5. If the redshift is less than
0.1, the de-absorbed spectrum will present the usual SED of an LBL
such as W Comae (\cite{Acciari08}). In the Fig.~\ref{fig.3}, we show
the de-absorbed SEDs under z=0.1 and the modeling. A smooth spectrum
can link the Fermi-LAT and VERITAS data and be explained with a SSC
model, in which $U_B=9.95 \times 10^{-5} erg \cdot cm^{-3}$ and
$U_e=2.31 \times 10^{-3} erg \cdot cm^{-3}$.

In fact the bulk motion of the emitting blob affects the observed
SED (e.g., \cite{Dermer95}; Georganopoulos et al. (2001)). The peak
frequencies are given by $\nu_s\approx 3.7 \times 10^6 \gamma
_{peak}^2 \delta B /(1+z)$ , $\nu_{SSC}\approx \frac{4}{3}\gamma
_{peak}^2 \nu_{s}$, and $\nu_{EC}\approx \frac{4}{3} \gamma
_{peak}^2 \nu_{ext} \Gamma \delta /(1+z)\label{nu_ec}$. We can see
that $\nu_{EC}$ would be larger than $\nu_{SSC}$ provided the blob
has larger $\delta$ or $\Gamma$ (Usually the viewing angle
$\theta\sim 1/\Gamma$ is assumed, $\delta\sim \Gamma$.) From the
ratio of peak luminosity, $\frac{L_{EC}}{L_{SSC}} \approx
\frac{L_{EC}^{'}}{L_{SSC}^{'}} =
\frac{\dot\gamma_{peak,EC}}{\dot\gamma_{peak,SSC}} \approx
\frac{U^{'}_{ext}}{U^{'}_{syn}} \approx \delta^4 \Gamma^2 \frac{\xi
L_{ext}}{L_s} \frac{R^2}{R^2_{ext}}$ ($U_{ext}$ is amplified by a
factor of $\Gamma^2$, see the Sikora et al. (1994) and Dermer
(1995)), where $\xi$ is the reproduce fraction of the $L_{ext}$, we
show that $\frac{L_{EC}}{L_{SSC}}$ is strongly affected by the bulk
motion of the blob. 
In the Fig. 1, using $U_B^{'}=4.6 \times 10^{-5} erg \cdot cm^{-3}$,
$U_e^{'}=1.67 \times 10^{-3} erg \cdot cm^{-3}$, and $U_{ext}^{'}
=3.5 \times 10^{-6} erg \cdot cm^{-3}$ we can model the SED.
$U_{ext}^{'}$ is obviously lower than $U_B^{'}$, however, the EC
luminosity is comparable with the synchrotron ones (see the Fig. 1).
In the observer frame, the beaming factor is different for EC
($\delta ^{4+2\alpha}$ (\cite{Dermer95}; Georganopoulos et al.
2001)), synchrotron and SSC emission ($\delta ^{3+ \alpha}$). The
difference of EC and synchrotron luminosity is reasonable by
considering their beaming factor.

The EC emission is less clear for the BL Lac objects. The lack of
strong emission lines and UV excesses suggests that the external
photon density $\xi L_{ext}$ is very low, while the Lorentz factor
of BL Lac objects is typically smaller than that of quasars
(\cite{piner08}). Their high energy emission strongly favors the SSC
mechanism over the EC mechanism. But, 3C 66A might be an exception
and reveal an existence of larger bulk velocity in the high energy
emissive region. Therefore, the high energy emission caused by EC
mechanism is likely observed in the IBL. This possibility needs
future Fermi-LAT and VERTAS observations for 3C 66A and a precise
redshift determination.

\begin{acknowledgements}
We thank the referee for a very helpful and constructive report
which helped to improve our manuscript substantially. We acknowledge
the financial supports from the National Natural Science Foundation
of China 10673028 and 10778702, and the National Basic Research
Program of China (973 Program 2009CB824800).
\end{acknowledgements}


%
%

%
%
%

%

\begin{table}
\begin{center}
\caption{Parameters for the SSC+EC or SSC model used to reproduce
the SED (Fig.~\ref{fig.1} and Fig.~\ref{fig.3}).\label{param}}
\vskip 12pt
\begin{tabular}{llll}
\hline
Parameters          &If z=0.444           &If z=0.1      \\
\hline
\hline
$L_{\rm inj}$ ($10^{41}$~erg~s$^{-1}$)          & $4.0$                         & $6.3$       \\
$\gamma_{\rm min}$                             & $2.5\times10^{4}$             & $1.5\times10^{4}$       \\
$\gamma_{\rm max}$                             & $8.0\times10^5$               & $8.0\times10^5$       \\
$s$                                            & $3.4$                         & $3.3$       \\
$R$ ($10^{16}$~cm)                              & $7.0$                         & $7.0$           \\
$\delta$                                       & $35$                          & $30$        \\
$B$ (Gauss)                                     & $0.034$                       & $0.05$       \\
$\nu_{ext}$ (Hz)                                & $2.5\times10^{15}$           & $--$      \\
$U_{ext}$ ($erg/cm^3$)                          & $3.5\times10^{-6}$            & $--$      \\
\hline
\end{tabular}
\end{center}
\end{table}

\end{document}